\documentclass[%
 reprint,
 amsmath,amssymb,
 aps,
]{revtex4-2}

\usepackage{graphicx}% Include figure files
\usepackage{overpic}
\usepackage{dcolumn}% Align table columns on decimal point
\usepackage{bm}% bold math
\usepackage{siunitx}%Unidades en el sistema internacional
\usepackage{ragged2e} % for \justifying

\usepackage{xcolor}
\usepackage{caption}
\usepackage{subcaption}
\usepackage{float} 
\usepackage[hidelinks]{hyperref}

\usepackage{appendix}

\begin{document}

%\preprint{APS/123-QED}

\title{Optimal array geometries for kinetic magnetism and Nagaoka polarons}
%Nagaoka polarons and their robustness under magnetic fields in finite-size arrays.\\
%Robust Nagoka polarons in optimized arrays}% Force line breaks with \\
%\thanks{A footnote to the article title}%

\author{N. Hernandez-Cepeda}
\email{nh047522@ohio.edu}
\author{Sergio E. Ulloa}
%\email{ulloa@ohio.edu}
\affiliation{
 Department of Physics and Astronomy and Nanoscale and Quantum Phenomena Institute, Ohio University, Athens,
Ohio 45701-2979, USA
}%
 %\altaffiliation{Instituto de Física, Universidad Autónoma de Puebla, Apartado Postal J-48, Puebla, Puebla 72570, Mexico}
 %Lines break automatically or can be forced with \\

%\collaboration{MUSO Collaboration}%\noaffiliation

\date{\today}% It is always \today, today,
             %  but any date may be explicitly specified

\begin{abstract}
Quantum dot (QD) platforms have enabled the direct observation of Nagaoka ferromagnetism (NFM) in small arrays and non-infinite interaction strength. However, optimizing the cluster connectivity characteristics that yield a ground state with maximal spin and their robustness against magnetic fields remains unexplored. Employing exact diagonalization of the Hubbard Hamiltonian, we find a connection between the existence of kinetic ferromagnetism and graph theory descriptions. Algebraic connectivity ($\lambda_2$) and Katz centrality (KC) are
shown to be related to the spin-correlation over the system. In square arrays, the onset of NFM is found to be $t_c/U \simeq\lambda_2^2$. In optimal cluster geometries, large $\lambda_2$ and low KC fluctuation per site are found to enhance $t_c/U$, extending the NFM phase while diminishing the strength of spin correlation clouds.  A perpendicular magnetic field    
introduces Aharonov-Bohm phases, and a critical flux for which NFM is destroyed. We further find that tuning the flux phase to $\pi$ results in a ground state that exhibits antiferromagnetic correlations (counter-Nagaoka state). Our results illustrate how NFM and polaron formation can be predicted from the array’s connectivity ($\lambda_2$ and KC), and how the introduction of flux results in the counterintuitive destruction of kinetic ferromagnetism in the system.

\end{abstract}

%\keywords{Suggested keywords}%Use showkeys class option if keyword
                              %display desired
\maketitle

%\tableofcontents

\section{Introduction}
Understanding the key ingredients that favor metallic ferromagnetism in materials is still a subject under intense investigation, especially as it refers to the role that quantum path interference plays for itinerant electrons. One unexpected form of magnetic ordering in itinerant electron systems was studied by Nagaoka using the Hubbard model \cite{nagaoka1966ferromagnetism}. He demonstrated that a ferromagnetic ground state emerges in certain two-dimensional lattices from interference of hole trajectories in a system doped away from half-filling, in the limit of infinite interactions. 

Nagaoka's result appears counterintuitive in light of two well-known results: 1) according to the Lieb-Mattis theorem, no ferromagnetic ground state can emerge in open one-dimensional lattices where only nearest-neighbor hoppings are considered \cite{PhysRev.125.164, mattis2003eigenvalues}; 2) a strongly interacting two-dimensional lattice with electrons at half-filling exhibits a ground state with antiferromagnetic (AFM) order given by the exchange energy $J \approx t^2/U$ \cite{RevModPhys.63.1, tasaki2020half}. 

%Under this perspective, it is remarkable that periodic boundary conditions (e.g., from a 4-site chain to a 4-site loop) and the motion of a single hole favor ferromagnetic order.

In the Nagaoka ferromagnetic (NFM) ground state the constructive interference of the paths taken by the hole in the system leads to equivalent spin configurations, favoring ferromagnetic alignment \cite{PhysRevB.40.9192}. NFM is characterized by $\textit{S}_{loop}$, the sign of the product of hopping amplitudes around the closed loop that encloses the smallest path taken by the hole \cite{PhysRevLett.95.087202, PhysRevB.95.195103}. If $\textit{S}_{loop} = +1 \, (-1)$, the array is formed by elementary loops of four (three) sites, and the motion of the hole leads to a ground state that favors NFM (AFM or counter-Nagaoka).
As a result of these observations,
the question of how connectivity affects the onset of NFM has been under discussion in the literature for different systems.

For instance, it was found that non-bipartite (triangular and kagome) lattices could host NFM for filling above half, whereas bipartite (honeycomb) lattices could do it for near half-filling \cite{hanisch1995ferromagnetism}. Tasaki \cite{ tasaki1998nagaoka} pointed out that by allowing electron hopping between any two sites, long-range ferromagnetism can arise at finite Coulomb interaction $U$, contrary to what was known about NFM (appearing in the $U \rightarrow \infty$ limit). This idea is particularly intriguing since it suggests that by changing site hoppings (connectivity of the nodes) it is possible to favor the emergence of ferromagnetic ground states. 

Another way to study the relationship between connectivity and magnetic ordering in the almost half-filled regime of interest is to consider frustrated hoppings \cite{PhysRevB.95.195103, PhysRevLett.112.187204}. These hoppings between next-nearest neighbors change a square lattice into one with triangular loops, changing the sign of $S_{loop}$, and a transition in the system from NFM to AFM. An alternative approach shows that connectivity and magnetic ordering can be tuned by including staggered magnetic fluxes in square and triangular lattices \cite{PhysRevLett.100.037202}. This leads to a similar transition from NFM to AFM ground states with one hole in the half-filled infinite-interaction system. Interestingly, it was recently shown that ferromagnetic clouds emerge around a doped hole (doped particle) in square (triangular) lattices in both the nearly half-filled $t$-$J$ and extended Hubbard models \cite{PhysRevB.64.024411,PhysRevA.110.L021303, PhysRevB.109.235128}. Related work in doped semiconductors provides complementary insights  \cite{PhysRevB.76.161202,PhysRevB.82.195117}. These results support the notion that connectivity plays a critical role in the emergence of ground states, and that it is possible to track the nature of the emerging magnetic ordering through spin correlation clouds in the array.

More recently, the experimental demonstration of NFM in finite-size arrays of QDs with different connectivities \cite{dehollain2020nagaoka} have motivated different theoretical studies 
on the conditions to host ferromagnetic ground states \cite{buterakos2019ferromagnetism, xavier2020onset, PhysRevB.110.245141}. Linear and ring geometries have been explored, finding different filling and connectivity conditions for the onset of different magnetic orderings. It has been shown that NFM occurs at large but finite values of the Hubbard interaction in square arrays and loops (see \cite{xavier2020onset} for a discussion of the appearance of ferromagnetism in loops). The critical hopping value below which NFM exists in large square arrays is found to decrease as a power law with increasing size of the array \cite{PhysRevB.110.245141}.

Experimental implementations of the Hubbard model and NFM candidate systems have been achieved in diverse platforms, including cold atoms, two-dimensional materials, and arrays of quantum dots (QDs). As mentioned, a first experimental probing of NFM was carried out in a 2$\times$2 array of QDs \cite{dehollain2020nagaoka}. Nagaoka polarons--ferromagnetic correlations surrounding a `doublon' (doubly-occupied site)--have been observed in triangular lattices of ultra-cold atoms \cite{lebrat2024observation, prichard2024directly}, while signatures of NFM were uncovered in the electronic magnetization of moire heterolayer materials \cite{ciorciaro2023kinetic}. These platforms have shown that NFM exists at finite interaction $U$ values, and that triangular lattices can also support NFM when a single electron is added above half-filling. These results reveal a close connection between NFM, lattice connectivity, and electron doping in different lattices. As several platforms can tailor the connectivity between sites in alternative geometries, away from square and triangular, it is appealing to identify the connectivity conditions required to attain the NFM regime at the smallest $U$ values and different hole doping levels. 

QD platforms offer unique advantages, given the high control over tunneling, on-site interaction, and filling factor per site \cite{barthelemy2013quantum}. These structures have enabled measurements of spin correlations \cite{mukhopadhyay20182}, control over individual electron spins \cite{PRXQuantum.2.030331}, exploration of quantum phase transitions in square arrays \cite{wang2022experimental}, electron number manipulation in
crossbar arrays \cite{borsoi2024shared}, and recent investigation of excitations in QD ladders \cite{PhysRevX.14.011048}. 

The role of connectivity in stabilizing or modifying NFM in finite-size arrays has resulted in a few open questions. In particular, one wonders whether ferromagnetic polarons can form in few-site clusters, and how they would be modified by the tunable connectivity, beyond the well-studied square and triangular geometries. Is there a connection between graph theory describing a QD cluster and the onset of NFM? Moreover, since it is possible to tune NFM by modifying $\textit{S}_{loop}$,  it is interesting to explore how robust NFM is under a magnetic flux through the cluster, as the resulting Aharonov-Bohm phases alter interference patterns of the hole (or extra electron) dynamics.

The remaining paper is organized as follows: in Sec.\ II, we discuss the Hubbard model and Nagaoka polarons in conventional square arrays. We introduce algebraic connectivity and Katz centrality, proposed in the context of graph theory, as the criteria to select the optimal finite-size array geometries. NFM occurs at a finite value of $t/U$, which can be enhanced by maximizing algebraic connectivity and Katz centrality per site. Nagaoka polarons are a signature of the NFM state, and we discuss their intensity dependence on connectivity for square of different size and for arrays with different geometry but same site number. In Sec.\ III, we present the effect of an Aharonov-Bohm phase on the selected geometries, finding that a magnetic field can surprisingly destroy a ferromagnetic phase. Finally, in Sec.\ IV, we summarize our findings on kinetic ferromagnetism in few-site arrays.

\section{Hubbard model in QD arrays}

\begin{figure}[h]
    \centering
    \begin{subfigure}[b]{0.45\textwidth}
        \begin{overpic}[width=\linewidth,grid=false]{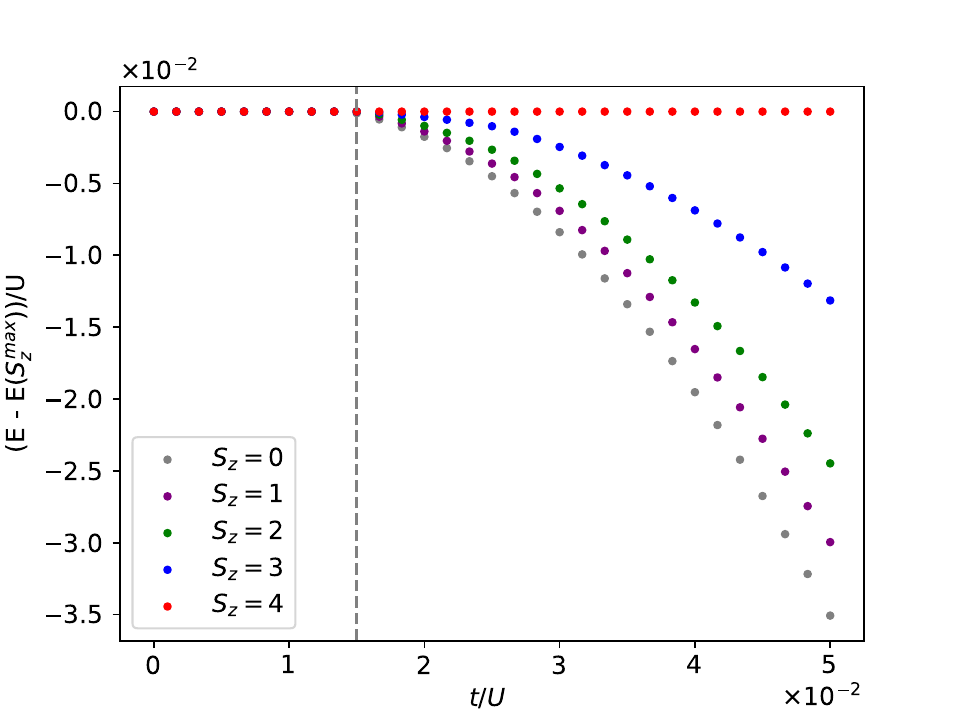}
            % coordinates: (x,y) in percent, (0,0)=bottom-left, (100,100)=top-right
            \put(2,70){\large{(a)}} % label in top-left
        \end{overpic}
        \centering 
        %\caption{ }
        \label{fig:fig_3x3a}

    \end{subfigure}
    \begin{subfigure}[b]{0.48\textwidth}
        \begin{overpic}[width=\linewidth,grid=false]{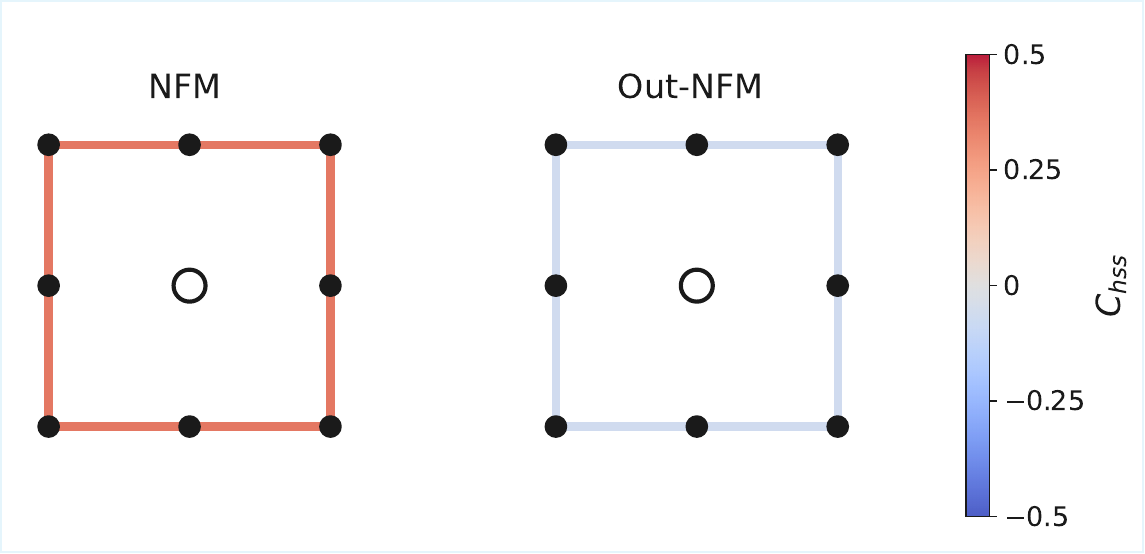}
            % coordinates: (x,y) in percent, (0,0)=bottom-left, (100,100)=top-right
            \put(5,45){\large{(b)}} % label in top-left
        \end{overpic}
        \centering 
        %\caption{}
        \label{fig:fig_3x3b}
    \end{subfigure}
    \centering
    \caption{\justifying {(a)} Lowest energy manifold in a $ 3 \times 3$ square array with 8 electrons (one less than half-filled) as a function of $t/U$. Transition to the NFM regime occurs for  $t/U < t_{c}/U \approx 0.0144$. (b) Hole-spin-spin correlator $C_{hss}$ in  ($t/U=0.01$), and beyond ($t/U=0.02$) the kinetic ferromagnetic regime, showing the relative spin alignments in the vicinity of the hole at central site (white circle).}
    \label{fig_3X3_array}
\end{figure}

The Hubbard model provides a framework for describing electrons tunneling between atomic orbitals localized on lattice sites. In this model, electrons hop between sites and experience a repulsive onsite interaction \cite{hubbard1963electron}. This description considers one orbital per site, assuming other orbitals do not contribute to the low-energy physics description of the system. Strong repulsion between electrons on the same site is important while the intersite interaction is neglected (typically much smaller \cite{fazekas1999lecture}). The corresponding Hamiltonian is given by
\begin{equation}\label{eq:1}
    \mathcal{H}=-  \sum_{\langle ij\rangle \alpha} t_{ij} c_{i, \alpha}^\dagger c_{j, \alpha} +U \sum_{i}n_{i,\uparrow}n_{i,\downarrow}.
\end{equation}
The hopping amplitude $t_{ij}$ represents the electron tunneling between nearest-neighbor sites \textit{i} and \textit{j} while preserving the spin $\alpha$, and \textit{U} is the on-site electron-electron interaction. The operators $c_{i, \alpha}^\dagger, c_{i, \alpha}$, and $n_{i, \alpha} = c_{i, \alpha}^\dagger c_{i, \alpha} $ are the creation, annihilation, and number operators, respectively, for a fermion on site $i$ and spin $\alpha$={↑, ↓}.
The eigenvalue spectrum can be obtained by exact diagonalization of the Hamiltonian, properly partitioned into blocks corresponding to different $ S_z$ manifolds, to take advantage of $S_z$ being a good quantum number. Eigenenergies for $ \pm S_z $ are the same, respecting time-reversal symmetry.

We now revisit the associated energetics and spin correlations in square clusters to set the stage for other geometries. Figure \ref{fig_3X3_array} shows the low-energy spectrum of a $3 \times 3$ array as a function of $t/U$, labeling states by the corresponding $S_{z} \geq 0$. 
NFM is known to appear in few-site arrays if they are connected in elemental loops of four sites for $t_{ij}>0$, as mentioned \cite{PhysRevB.40.9192, PhysRevB.98.180101}. 
NFM arises when the total spin in the ground state is maximum, $S_{max}= N/2$, with degeneracy given by $2S_{max} +1$; $N$ is the number of electrons in the array, one less than the number of sites.  As described by Nagaoka \cite{nagaoka1966ferromagnetism,tasaki1998nagaoka}, the NFM regime is reached at infinite $U$, seen to be the case for $t/U<t_c/U$ in Fig.\ref{fig_3X3_array}, as the lowest state of each $S_z$ subspace is degenerate with each other. After the transition ($t/U>t_c/U$), the degeneracy is broken, as the lowest state in each $S_z$ subspace relates to different total spin. Notice that the ground state for $t/U>t_c/U$ in this case corresponds to the array singlet with $S=0$.

\subsection{Nagaoka polarons}

An intuitive picture of the onset of the NFM regime and the role that the hole plays have been described before as a Nagaoka polaron \cite{PhysRevB.64.024411,PhysRevB.109.235128, lebrat2024observation, prichard2024directly}. This can be illustrated by the three-point correlator that determines the relative spin alignment in the hole's vicinity
\begin{equation}\label{eq:2}
    C_{hss}^{m}= \frac{\langle h_{k} \vec{S_{i}}\cdot \vec{S_{j}} \rangle _{m}}{ \langle h_{k} \rangle_m },
\end{equation}
where $h_{k} = (1-n_{k\uparrow})(1-n_{k\downarrow})$ is the hole operator at site $k$, $\vec{S_{i}}$ is the spin operator at the $i$-th site, and $ \langle \mathcal{O} \rangle _{m} = \langle \psi^{S_{z} = m} |\mathcal{O}| \psi^{S_{z} = m} \rangle$ is the expectation value of the $\mathcal{O}$ operator over the state with $S_{z}=m$.  The denominator in Eq.\ \ref{eq:2}, the expected value of the hole at site $k$, is introduced to highlight the spin correlation between $i$ and $j$ sites in each $S_z$ subspace. $C_{hss}^{m}$ yields the value of the intersite spin correlation once the hole is fixed at a particular site on the array. For completely aligned spins, one expects $C_{hss}^{m} = 1/4$ (ferromagnetic correlation), while for antiferromagnetically correlated spins, this value is $-3/4$.

Figure \ref{fig_3X3_array}b depicts $C_{hss} =\sum_{m} C_{hss}^{S_{z} = m}$ within and outside the NFM regime. In the NFM regime, $t/U < t_c/U$, $C_{hss}$ shows a clear ferromagnetic correlation cloud surrounding the hole position, as shown in the left map. The correlation remains constant throughout the region $t/U < t_c/U$ (with a value here of $C_{hss}=0.33$).  After the transition (right map), $C_{hss} = C_{hss}^{S_{z} = 0}$, since the ground state is unique and given by the $S_z=0$ block, and the cloud has an intensity of $C_{hss} \simeq -0.06$, indicating that AFM correlations are favored as $t/U$ increases. Thus, as discussed in experiments and large array theoretical studies, the presence of a hole allows the formation of ferromagnetic correlations between neighbors, the Nagaoka polaron, evident even in small arrays. These polarons have been illustrated using DMRG in the ground state of the $t$-$J$ model \cite{PhysRevB.64.024411}, as well as in an extended Hubbard model with occupation-dependent hoppings \cite{PhysRevA.110.L021303, PhysRevB.109.235128}.

\subsection{NFM and graph theory}

The degree of connectivity and parity of closed loops in a given structure have been shown to influence the onset of NFM \cite{PhysRevB.98.180101, PhysRevB.110.245141,s2025shapingmagneticorderlocal}. This behavior suggests a possible connection between the properties of an array and its graph theory representation. With this in mind, we construct the Laplacian matrix of an array, $\mathcal{L}= D - A$, used in graph theory to describe these structures. $\mathcal{L}$ is given in terms of the adjacency matrix, $A$, which describes how the sites of the graph/array are connected, and the degree matrix $D$ that counts the connections of each graph node. The eigenvalue spectrum of the Laplacian matrix describes different connectivity conditions of the array. The graph is connected if the first eigenvalue is unique and equal to zero, while the second eigenvalue is greater than zero, so that $\lambda_1 = 0$ and $\lambda_{2}>0$; similarly, the graph is bipartite if
the largest eigenvalue of the normalized Laplacian is 2 \cite{fiedler1989laplacian, chung1997spectral}. The $\lambda_{2}$ eigenvalue (Fiedler value or algebraic connectivity) determines how well-connected a graph is, as it represents better communication between nodes, describing synchronization or diffusion processes in networks \cite{mohar1991laplacian}. In our case, it is the hole that delocalizes as the connectivity increases, allowing for multiple path interference. This suggests that different finite-size systems can consistently exhibit a ferromagnetic ground state with one hole away from half filling for finite values of $t/U$, and that it may be related to the $\lambda_2$ values.

In square arrays, the largest $\lambda_2$ occurs for the $2\times2$ array ($\lambda_2=2$), and progressively decreases for larger arrays.  This dependence on array size is shown in Fig.\ \ref{fig:lambda2} in App.\  \ref{Appendix}, where we see that $\lambda_2 \simeq L^{-2}$ for square arrays of size $L \times L$.  Interestingly, a different power law with size is seen for the critical value $t_c/U$ for which the NFM is achieved. One finds $t_c/U \simeq L^{-4}$, as reported recently \cite{PhysRevB.110.245141}. This rapid decrease of $t_c$ with size can be seen as the natural increase in the number of interfering paths available to 
the doped hole/electron as the area of the array increases. The drop in algebraic connectivity can be understood from the Laplacian eigenvalue expression for a square grid, $\lambda = 4-2\cos(n\pi/L)- 2\cos(m\pi/L')$, obtained as the cartesian product of two path graphs of $L$ and $L'$ nodes, where $n, m$ are integer indices that quantize the eigenvalues in the $L$ and $L'$ direction, respectively
\cite{fiedler1989laplacian}. When $L = L'$ and $(n,m) = (1, 0) = (0, 1)$ the eigenvalue expression describes the second lowest eigenvalue $\lambda_2$ for a finite square grid. As $L \rightarrow \infty$, $\lambda_2 \simeq \pi^2 /L^2$, as we see in Fig. \ref{fig:lambda2} in App.\ \ref{Appendix}.

The dependence $t_c/U \simeq L^{-4} \simeq \lambda_2^2$ for square arrays has a form of inverse area squared law, and the intensity of a signal spreading out is proportional to the square of the distance it travels. From this perspective, it seems that the coherent spreading of the hole, which gives rise to NFM decays as $L^{-2}$ in each direction of the square array. Thus, as the system size increases, the hole can effectively move less on the array (low $\lambda_2$), and NFM is only possible at infinite interactions.

Note 
$\lambda_2$ is not the only connectivity parameter that helps us explain the NFM robustness in small-sized arrays. It is natural to expect that the spatial distribution of the hole, $P_{h_{i}} = \langle  h_{i} \rangle$, would be highly dependent on how well-connected with its surroundings is a site. Interestingly, we find that this probability can be described by the Katz centrality (KC) vector $\vec{x} = ( \mathbf{I} -\alpha A)^{-1} \mathbf{1}$, where $\mathbf{I}$ is the 
identity matrix,  $\mathbf{1}$ the vector with all entries equal to 1, and $\alpha$ and $A$ are the attenuation factor and adjacency matrix, respectively \cite{networks}. KC describes the importance of a site in an array ($x_i$) by determining the number of direct and indirect connections it has through $A$, as  well as how well connected these neighbors are ($\alpha$ penalizes connections with distant neighbors). The probability of finding the hole in a particular site increases with its KC value. We will illustrate below how the KC values of different arrays help us understand the hole propagation in a QD array.

\subsection{Optimal QD array geometries}

\begin{table}
    \captionsetup{justification=raggedright, singlelinecheck=false}
    \centering
    \begin{tabular}{llcllccl}
    \hline\hline
     Array & & Graph & & $t_{c}/U$ & & $\lambda_2$  \\
     \hline
     L  & & \includegraphics[width=0.4in]{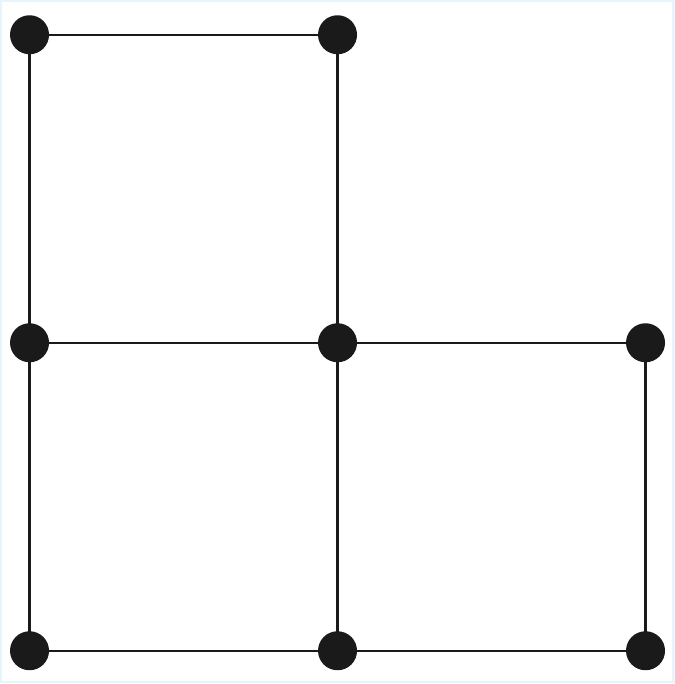}  & & 0.010 & &0.75 \\
     2$\times$4 Ladder  & & \includegraphics[width=0.6in]{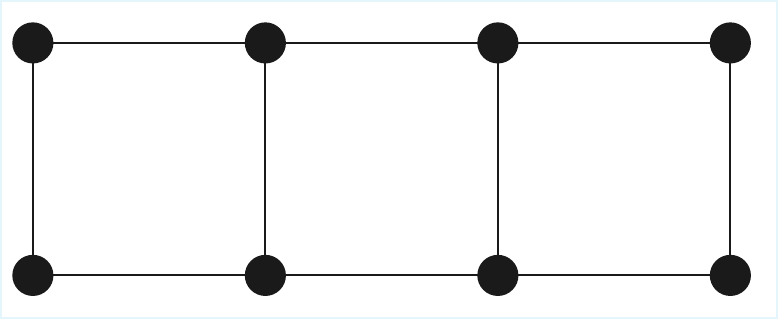}  & & 0.012  & &0.59  \\
     Kite  & & \includegraphics[width=0.6in]{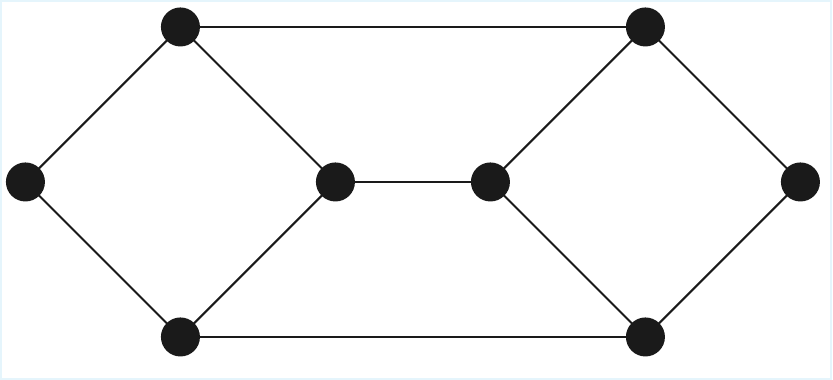}   & & 0.016  & &1.09 \\
     2-Square  & & \includegraphics[width=0.4in]{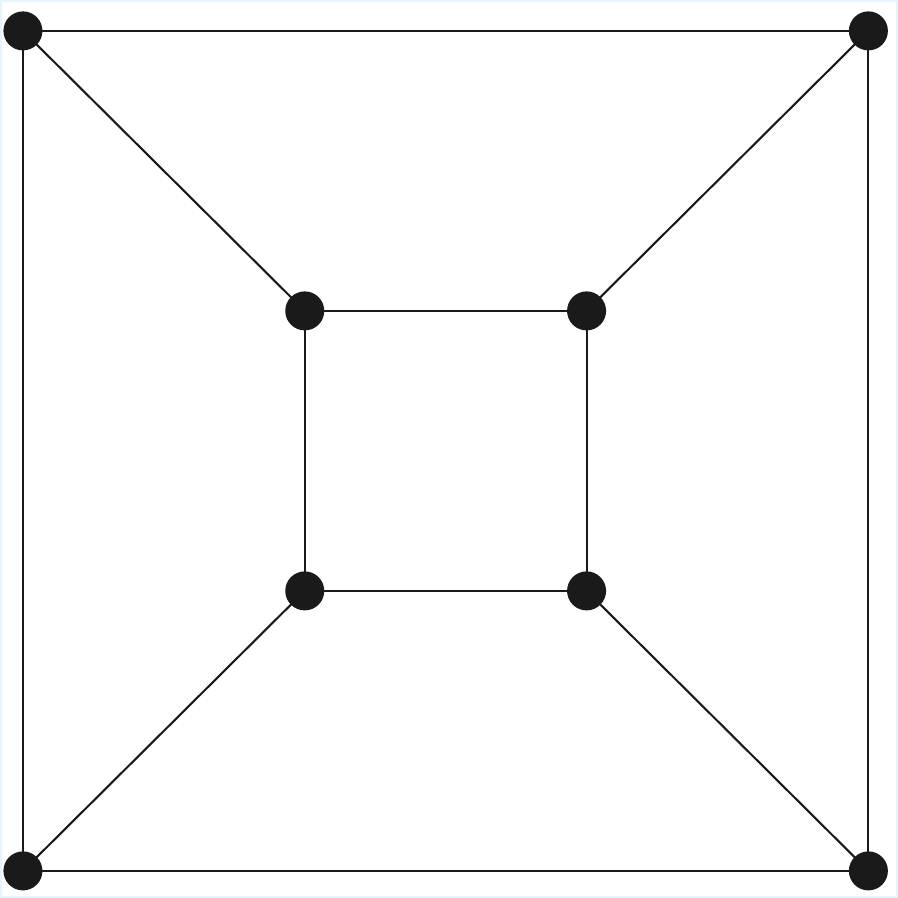}   & & 0.026  & &2 \\
     \hline\hline
    \end{tabular}
    \caption{Proposed QD arrays with 8 sites: Different columns show array label, graph, transition point ($t_c/U)$, and algebraic connectivity $\lambda_2$.}
    \label{tab:geometries_8M}
\end{table}

The connectivity considerations described in the previous section suggest an approach to optimize the geometry of the QD cluster to maximize the value of $t_c/U$ that allows for NFM in the system, since $t_c/U \simeq \lambda_2 ^2$. We focus on bipartite graphs that maximize both $\lambda_2$ and KC per site, and yet contain elemental loops with $\textit{S}_{loop}=1$. We focus on clusters with eight sites (or QDs) that are perhaps less challenging to implement experimentally. The analysis can be carried out for other array sizes and arrangements (see App.\ \ref{AppendixB}). Table \ref{tab:geometries_8M} presents the proposed geometries, with their respective $t_c/U $ transition points, as well as $\lambda_2$ values. The resulting $t_c$ and $\lambda_2$ values 
change in parallel from structure to structure (although not perfectly monotonic). Moreover, KC offers a good insight: by keeping the KC value per site as large and uniform as possible, the hole path interference is expected to survive for larger $t_c/U$. We have kept the number of sites constant to control for the area scaling seen in the square arrays in the previous section.  We anticipate that the relation between larger $\lambda_2$, homogeneous KC, and larger $t_c$ for a given number of sites would be a general behavior in QD clusters.  We have explicitly verified this for clusters of ten sites (see App.\ \ref{AppendixB}).

As the number of 3-link sites increases while suppressing sites with bulk connectivity (four-links), $t_{c}/U$ is seen to increase by nearly a factor of $\sim 3$ in Table \ref{tab:geometries_8M}. Bulk sites have a higher KC since they have more nearest neighbors, which suggests that the hole would be distributed more easily in the array. However, we find the opposite: high KC of a site in an array results in a higher probability of finding the hole in that site. As the hole is not well spread out over the array, one finds the ratio $t_c/U$ to decrease. 

\begin{figure}[h]
    \includegraphics[width=1.0\linewidth]{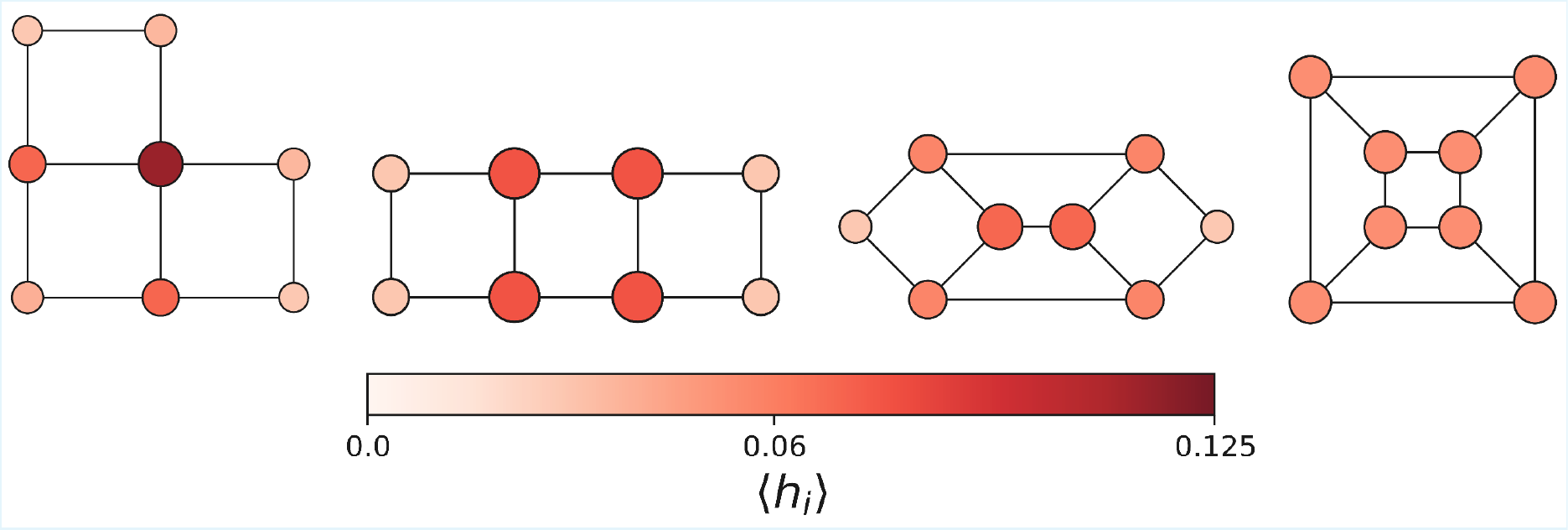}
    \centering
    \caption{\justifying Hole distribution per site in the clusters of Table \ref{tab:geometries_8M} at $t<t_c$. The color of each circle conveys $\langle  h_{i} \rangle$, while its diameter is proportional to the Katz centrality (KC) of the site. Notice that in the L-array, the hole is mostly concentrated in the central bulk-site (larger diameter), whereas in the 2-Square array (right-most graph), the hole is distributed uniformly (all sites have the same diameter). }
    \label{fig:Nagaoka_charge_8M}
\end{figure}

As a consequence, optimizing the geometry that hosts NFM with large $t_c/U$ value involves considering not only $\lambda_2$, but also the KC vector. This is how we can understand the non-monotonic trend in Table \ref{tab:geometries_8M}: Even though $\lambda_2({\rm L\, array})> \lambda_2(2\times4 {\rm \, array})$, we find  that $t_c({\rm L})< t_c(2\times4)$, as the L array has a more uniform KC vector. The Kite and 2-Square arrays in Table \ref{tab:geometries_8M} have both a higher algebraic connectivity and a more uniform KC vector, indicating the hole is more evenly distributed over the array (see Fig.\ \ref{fig:Nagaoka_charge_8M}).

We can conclude that arrays with a high variability of KC value per site suppress the effective hole's coherent motion that gives rise to kinetic ferromagnetism, and the optimal geometries reduce or totally avoid KC variability, while increasing $\lambda_2$.

The low-energy manifold for these clusters as a function of $t/U$ follows an analogous behavior to the one seen in Figure \ref{fig_3X3_array}a.
Notice that here $S_{max}= 7/2$ with an eight-fold degeneracy, and $t_c/U$ changes according to Table \ref{tab:geometries_8M}. When the system is outside of NFM regime ($t>t_c$), the ground state corresponds to $S=1/2$.

\subsection{Nagaoka polarons in optimal geometries}

\begin{figure}[h]
    \includegraphics[width=1.0\linewidth]{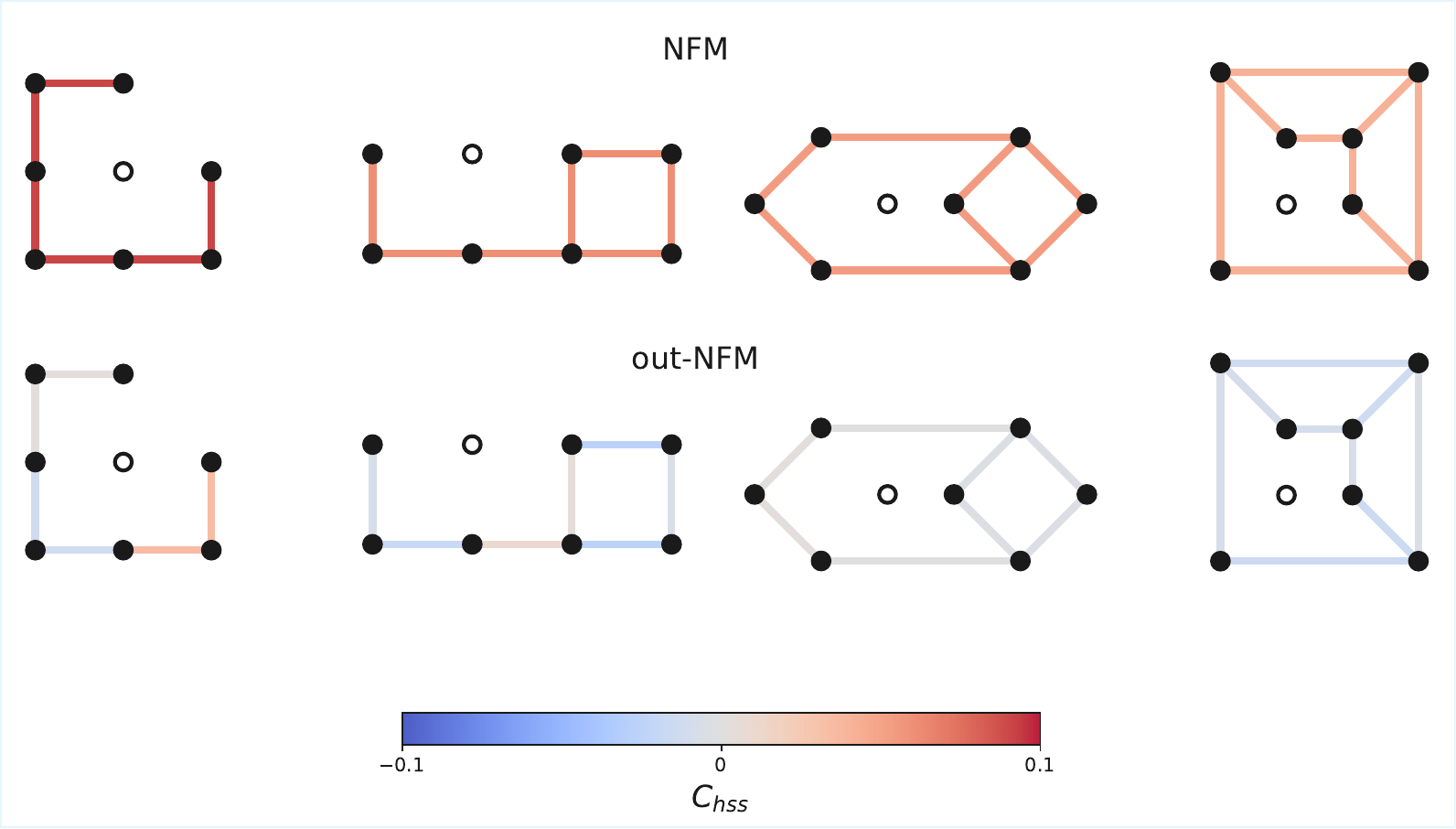}
    \centering
    \caption{\justifying Nagaoka polarons in clusters with different connectivities, in and out of the kinetic ferromagnetic regime. $\tilde C_{hss}$ determines the total spin-spin correlation in the vicinity of the hole (white dot). In the Nagaoka regime, upper row, $\tilde C_{hss}$ is shown for $t<t_c$ 
    to explore the signature ferromagnetic correlations; notice its higher value for the L-array (left-most). The signature ferromagnetic correlations vanish throughout for values of $t>t_c$,  as shown in the lower row just outside the NFM regime (at $ t/U = 0.011, 0.013, 0.018, 0.027$ from left to right). }
    \label{fig:Nagaoka_polarons_8M}
\end{figure}

As we aim to determine the role of geometry in the formation of Nagaoka polarons, we explore the $C_{hss}$ correlator for the geometries proposed in Table \ref{tab:geometries_8M}.
Figure \ref{fig:Nagaoka_polarons_8M} shows the corresponding hole-spin-spin correlator for the optimal clusters $\tilde C_{hss} = \sum_{m} C_{hss}^{S_{z} = m} \langle h_{k} \rangle$, keeping the number of sites (8) and electrons (7) constant. 
Notice that this quantity eliminates the hole normalization factor in Eq.\
\ref{eq:2}, in order to highlight the correlation strength in the different geometries.

In the upper row of Fig.\ \ref{fig:Nagaoka_polarons_8M}, we notice the clear ferromagnetic cloud surrounding the hole, illustrating the appearance of the Nagaoka polaron in these structures. The strongest spin correlations are visible in the L-geometry that allows the hole to connect to four sites. In the lower row, the Nagaoka polaron is destroyed when $t/U$ moves past the transition point. Notice that NFM correlations appear in sectors that contain elemental loops with four links, while each site has typically only 3 links. Although ferromagnetic clouds have been clearly seen in large square and triangular lattices \cite{sharma2025instabilitynagaokastatequantum, pereira2025kinetic}, it is interesting that they survive in few-site arrays and for larger critical $t_c/U$ values if the connectivity is properly designed. 

Notice that although the larger $t_c/U$ is achieved in the 2-Square array, the NFM correlations for $ t/U<t_c/U$ appear slightly weaker across bonds in Fig.\ \ref{fig:Nagaoka_polarons_8M}, associated with the evenly localized hole seen in Fig.\  \ref{fig:Nagaoka_charge_8M}.
These results provide insight into the relationship between algebraic connectivity and the KC vector, as well as the onset of kinetic ferromagnetism and the corresponding appearance of Nagaoka polarons.

While the connectivity of a site/QD to its neighbors can be turned on and off \cite{dehollain2020nagaoka}, affecting the appearance of NFM, 
it is interesting to explore if a gradual change may affect the appearance of NFM and polarons.  One possibility is provided by introducing a magnetic flux in the system, as we discuss in the following section.

\section{Tuning kinetic ferromagnetism with magnetic flux}

As discussed previously, NFM arises from the hole's motion in the elemental loop of four sites in a bipartite lattice. Thus, the question arises of how to affect this motion and, therefore, the constructive interference that makes NFM and Nagaoka polarons possible. 

An intuitive possibility with experimental relevance is to add an Aharonov-Bohm phase $\phi$ induced by a magnetic flux per plaquette in order to disturb the constructive interference phenomenon. We capture this by incorporating a Peierls phase ($\theta_{ij})$ in the hoppings of Eq.\ (\ref{eq:1}) as follows:

\begin{equation}\label{eq:3}
   \theta_{ij}= \frac{2\pi}{ \Phi_{0}} \int_{i}^{j} d\vec{r}\cdot \vec{A} (\vec{r} ),
\end{equation}
so that the total phase around the elemental loop around four sites is given by $\phi = \sum_{\rm loop} \theta_{ij}$, where
$\Phi_{0}=e/hc$ is the magnetic flux quantum, and $\vec{A}$ is the magnetic vector potential that produces a $B \hat{z}$ field perpendicular to the plane of the array, so that the flux per plaquette $\Phi = B \mathcal{A}$, with $\mathcal{A}$ being the area of the elemental loop.

\begin{figure}[h]
    \centering
    \begin{subfigure}[b]{0.46\textwidth}    
        \begin{overpic}[width=\linewidth,grid=false]{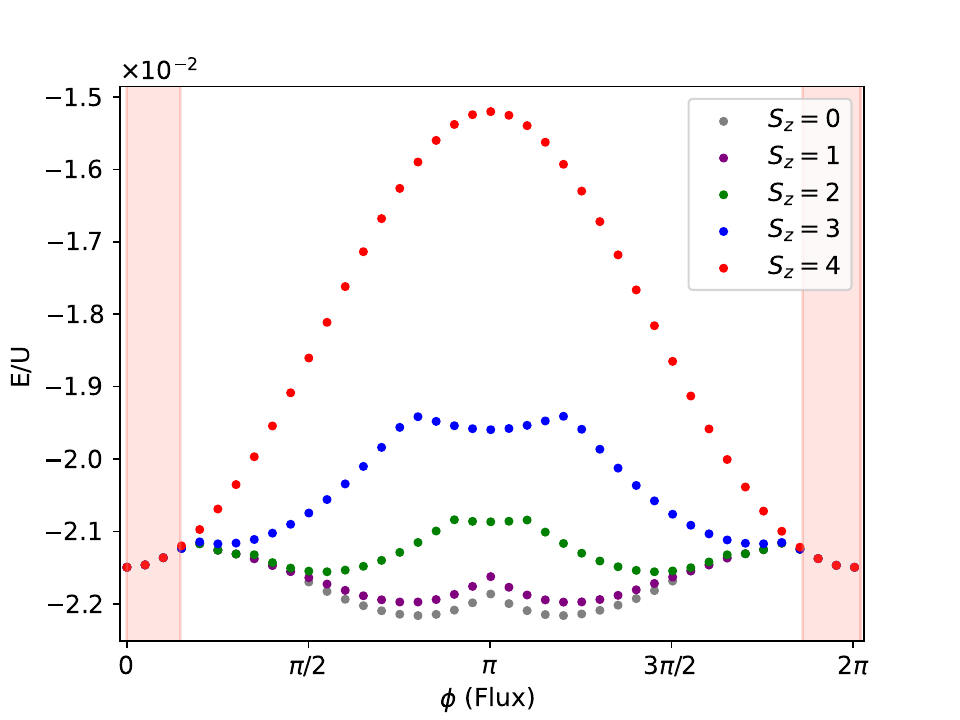}
            % coordinates: (x,y) in percent, (0,0)=bottom-left, (100,100)=top-right
            \put(2,70){\textbf{(a)}} % label in top-left
        \end{overpic}          
        %\includegraphics[width=\textwidth]{E_flux_B3_3X3.pdf}
        %\caption{ }
        \label{fig:fig_E_Bfield}
    \end{subfigure}
    \hfill
    \begin{subfigure}[b]{0.48\textwidth}
        \begin{overpic}[width=\linewidth,grid=false]{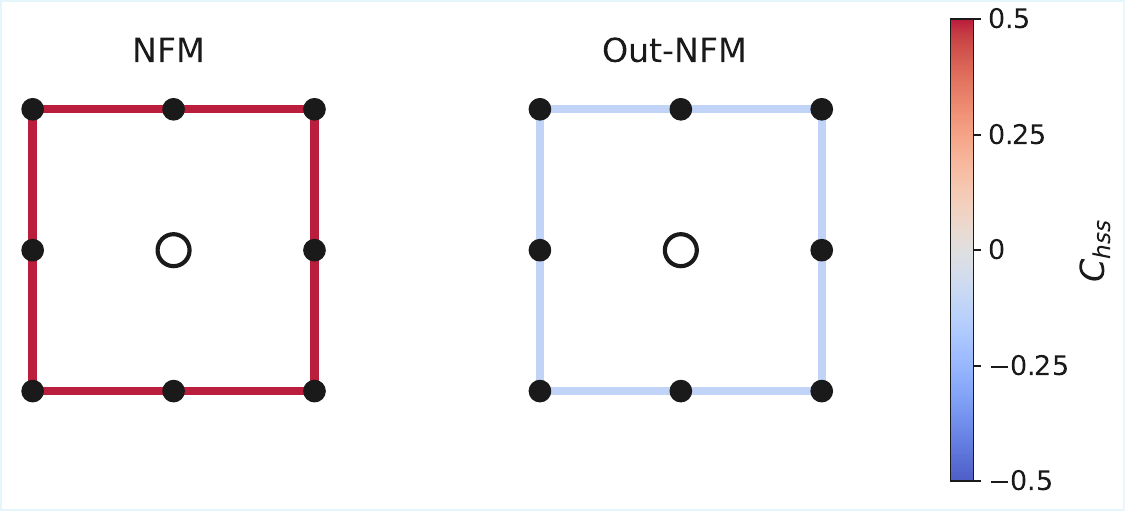}
            % coordinates: (x,y) in percent, (0,0)=bottom-left, (100,100)=top-right
            \put(2,46){\textbf{(b)}} % label in top-left
        \end{overpic}
        %\includegraphics[width=\textwidth]{PhiC_tc_Nagaoka_3X3_1.pdf}
        %\caption{}
        \label{fig: fig_Phi_Flux}
    \end{subfigure}
    \centering \caption{\justifying (a) Lowest energy levels in $3 \times 3$ array as a function of flux through elementary loop for the $S_{z} \geq 0$ manifold ($S_z<0$ subspaces have the same energy). The degeneracy is lifted for $|\phi|> \phi_c$ (mod $2 \pi$) and the system abandons the ferromagnetic ground state as different $S_z$ states split. (b) Hole-spin-spin $C_{hss}$ correlator at different fluxes. For $ |\phi| \leq \phi_c$ (pink shading region in (a)), the system exhibits a ferromagnetic cloud around the hole (white dot), while for $\phi \simeq\pi$ the ground state exhibits AFM correlations for any value of $t/U$.}
    \label{fig:fig_3X3_flux}
\end{figure}

\subsection{Square arrays}

Figure \ref{fig:fig_3X3_flux}a shows the effect on the energy spectrum of magnetic flux through a $3 \times 3$ array (we neglect a Zeeman contribution, assuming a near-zero $g$ factor). The ground states of each $S_z\geq 0$ manifold are plotted for a particular $t/U < t_c/U$, so that the system is in the NFM regime in the absence of flux. For small flux values, the full 9-fold degeneracy of the ground state is preserved (figure shows only $S_z \geq 0$ states, as the energies for $S_z<0$ are the same), indicating the system is still in the ferromagnetic phase. However, once the magnetic flux reaches a critical value, $\phi_{c}$, the system transitions to a low-spin state, destroying the ferromagnetic character of the ground state.

Once the system is out of the ferromagnetic regime, the Aharonov-Bohm phase produces an increasing energy arrangement of states from lower to larger spin projection. This energy splitting according to the $S_z$ subspace is an interesting result of path interference in the many-body state manifold.

To understand how NFM vanishes at a particular $\phi_c$, we explore the spin correlation behavior around a hole for the ground state, using Eq.\ (\ref{eq:2}). As depicted in Fig.\ \ref{fig:fig_3X3_flux}b (left), $C_{hss}$ is positive for $\phi < \phi_c$ 
%than $C_{hss}$ in Figure \ref{fig_3X3_array}b when $\phi = \phi_c$ and $t/U \le t_c/U$. For $\phi < \phi_c$, the correlation 
and has the same value as when no magnetic flux is present.  Interestingly, right at $\phi=\phi_c$ (and $t < t_c$), $C_{hss}$ is slightly larger than in Fig.\ \ref{fig_3X3_array} (not shown).
The strength of the Nagaoka polaron remains a fingerprint of the ferromagnetic phase, even in the presence of magnetic flux. 

%confirming that delocalization of the hole is what triggers the transition out of the ferromagnetic phase. \su{TRUE?}

Once the system leaves NFM (Fig.\ \ref{fig:fig_3X3_flux}b, right), the ground state is in the $S_z =0$ subspace, and $C_{hss}$ ($=-0.12$) shows clear AFM correlations at $\phi = \pi$. This behavior is expected, as $S_{loop} = -1$ for this flux value.

We have compared ground state energy as a function of $t/U$ for the array system under different conditions, including the well-known AFM at half-filling, 
and three different regimes for one hole away from half-filling: NFM ($S_{loop}=+1$), counter-Nagaoka ($S_{loop}=-1$ \footnote{Frustrated hoppings in the square array are introduced by allowing hopping between next-nearest-neighbors}), and $\pi$-AFM (magnetic flux equal to $\pi$ in the otherwise NFM case).  We find that the last configuration is more stable (has lower energy per particle) than both NFM and regular AFM, but less stable than counter-Nagaoka.
We also compare the AFM correlation cloud obtained in counter-Nagaoka systems \cite{Note1} with that in the square array with magnetic flux. As expected, we find that the $\pi$-AFM phase induced by the flux is fully equivalent to that generated in the counter-Nagaoka regime at the same $t/U$ value.
Thus, by tuning the magnetic field, we can controllably destroy the NFM phase and induce an AFM correlated phase that is more favorable energetically.

Figure \ref{fig: fig_Phi_Flux} shows that the critical flux $\phi_{c}$ is a linear function of $t/U$ for the $3 \times 3$ array. There is a critical flux in the limit of infinite interactions ($\phi_c \simeq \pi/4$) for which the ground state is no longer ferromagnetic. The behavior of the energy manifold with flux, and of $\phi_c$ with hopping, follows similar trends in other square ($ L \times L$) arrays.

\begin{figure}
    \centering
    \includegraphics[width=1.0\linewidth]{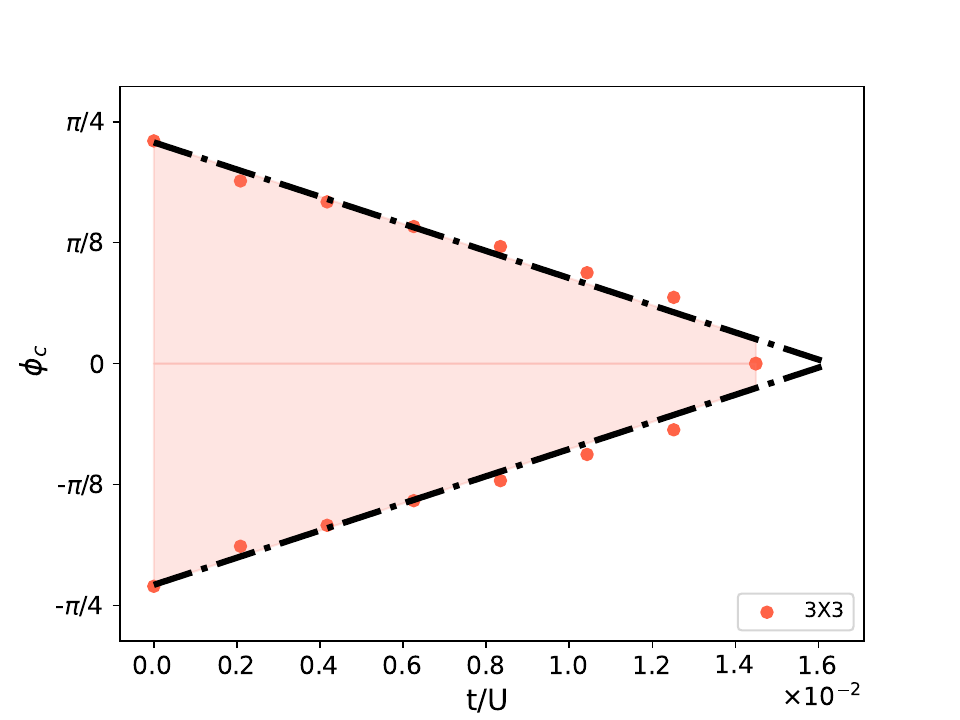}
    \caption{\justifying Critical flux as a function of $t/U$ in a 3$\times$3 array with one-hole away from half-filling. Dashed lines show the fit $\phi_c \simeq \pm (44.1 \, t/U - 0.23\pi)$ }
    \label{fig: fig_Phi_Flux}
\end{figure}

\begin{figure}
    \centering
    \includegraphics[width=1.0\linewidth]{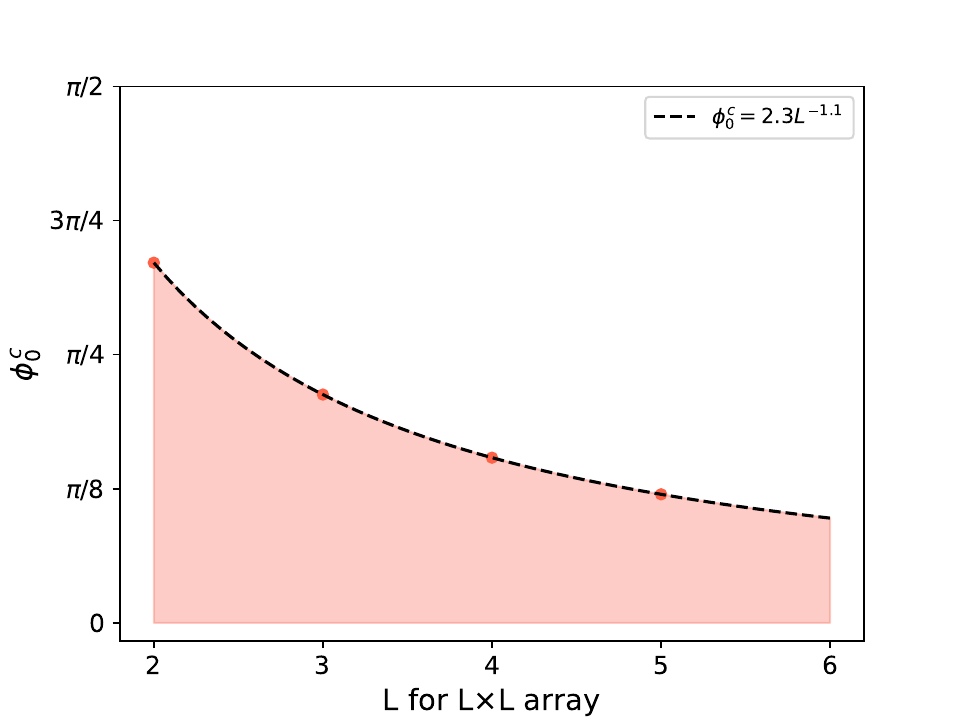}
    \caption{\justifying Critical flux as a function of size $L$ for square arrays in the limit \textit{U} $\rightarrow \infty$. }
    \label{fig:Phic}
\end{figure}

The size dependence of $\phi_c$ is shown in Fig.\ \ref{fig:Phic} for $ L \times L$ arrays in the limit of infinite interactions ($t/U \rightarrow0$). We see that when the net phase accumulated by the hole around a loop exceeds the threshold $\phi_{0}^{c} \, L \approx 2.3 $, it is sufficiently large to destabilize the kinetic ferromagnetic ground state. In the infinite 2D square lattice, the critical phase is zero, as expected from the vanishing $t_c/U$ discussed above \cite{PhysRevB.110.245141}.

This behavior reflects the interplay between the hopping phase acquired by the hole due to the applied flux and the constraints produced by correlations. As the system grows, the algebraic connectivity decreases, making the hole's constructive interference sensitive to even smaller magnetic fluxes. These more easily disrupt the constructive interference responsible for stabilizing the kinetic ferromagnetic phase.

\subsection{ Optimal QD arrays geometries in magnetic flux}

\begin{figure}[h]
    \centering
    \begin{subfigure}[b]{0.46\textwidth}    
        \begin{overpic}[width=\linewidth,grid=false]{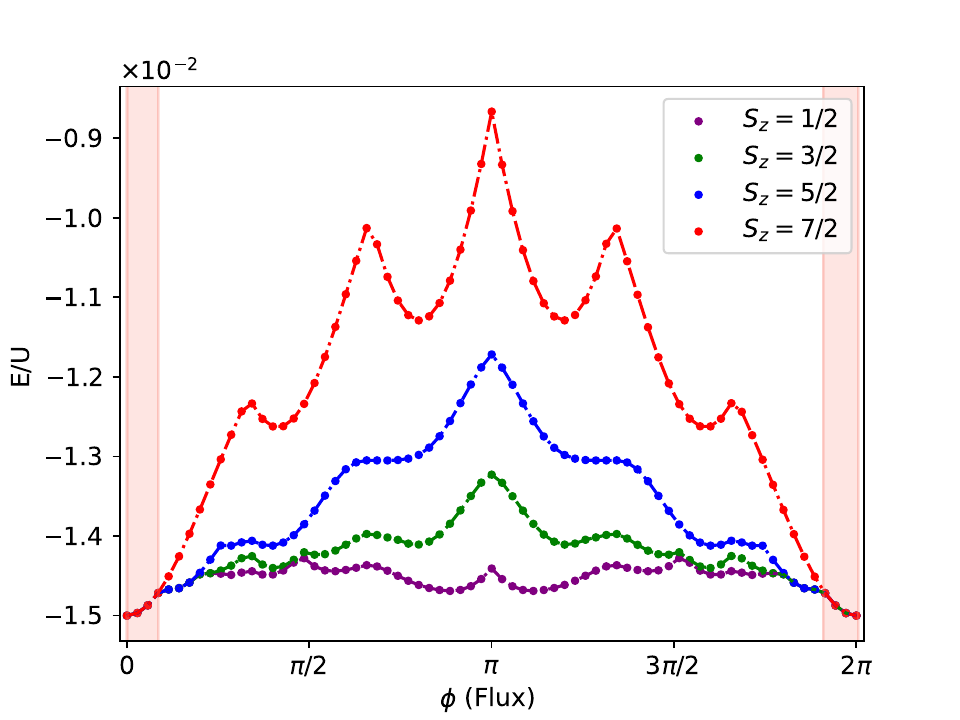}
            % coordinates: (x,y) in percent, (0,0)=bottom-left, (100,100)=top-right
            \put(2,70){\textbf{(a)}} % label in top-left
        \end{overpic}          
        %\includegraphics[width=\textwidth]{E_flux_B3_3X3.pdf}
        %\caption{ }
        \label{fig:E_phi_2square}
    \end{subfigure}
    \hfill
    \begin{subfigure}[b]{0.48\textwidth}
        \begin{overpic}[width=\linewidth,grid=false]{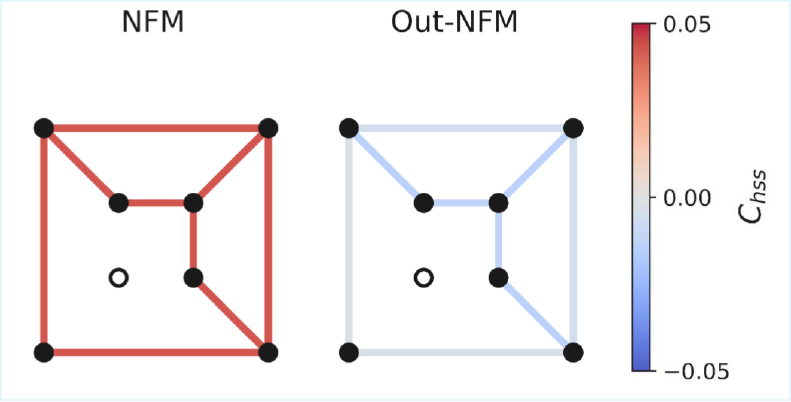}
            % coordinates: (x,y) in percent, (0,0)=bottom-left, (100,100)=top-right
            \put(2,48){\textbf{(b)}} % label in top-left
        \end{overpic}
        %\includegraphics[width=\textwidth]{PhiC_tc_Nagaoka_3X3_1.pdf}
        %\caption{}
        \label{fig: Chss_flux_2sq}
    \end{subfigure}
    \centering \caption{\justifying (a) Energy of lowest state for each $S_z$ manifold as function of flux for the 2-Square geometry (largest $\lambda_2$ in Table \ref{tab:geometries_8M}). (b) $C_{hss}$ correlator at different fluxes. For $0 <\phi \leq \phi_c$ , the system exhibits ferromagnetic correlations around the hole (white dot) with a peak at $\phi= \phi_c$ (left), while for $\phi=\pi$ results in AFM correlations for any value of $t/U$ (right).}
    \label{fig:fig_2square_flux}
\end{figure}

The $S_z$ manifold energy curves as a function of flux $\phi$ for the proposed geometries in Table \ref{tab:geometries_8M} have a qualitatively similar behavior to that in Fig.\ \ref{fig:fig_3X3_flux}a, with some important differences. In this section, we present the behavior for the 2-Square geometry, as it is the array to host NFM for larger $t_c/U$. As shown in Fig.\ \ref{fig:fig_2square_flux}a, the energy curve oscillates with flux (indicating more frequent ground state transitions), although the main characteristics remain: the NFM phase is present for low $\phi$, followed by splitting of the $S_z$ manifold for $\phi > \phi_c$, with the lowest-$S$ ground state (1/2 here). Figure \ref{fig:fig_2square_flux}b exhibits the expected Nagaoka polaron for $\phi \leq \phi_c$, and AFM spin clouds for $\phi = \pi$. 
It is interesting to note that the
hole distribution per site in this geometry exhibits a slightly higher concentration of the hole in the inner square than in the outer sites as the flux reaches $\phi_c$.
This is in slight contrast to the hole distribution in Fig.\ \ref{fig:Nagaoka_charge_8M}, where $ \langle h_{i} \rangle$ is the same for every site and all $t/U \le t_c/U  $. 
These correlation maps demonstrate that magnetic ordering can be tuned by the magnetic flux regardless of the geometry. From this perspective, we conclude that Nagaoka polarons can be created or destroyed by tuning the intensity of the applied magnetic field, while keeping the geometry of the QD array fixed.

The small $\phi_0^c $ and slope values in the 2-Square geometry suggest that even a weak magnetic flux would tend to reduce the extension of the hole over the array, suppressing the hole's interference of multiple paths. It appears that the same $\langle h_{i} \rangle$ per site, which is one of the features that makes NFM more robust in this cluster when varying $t/U$ at zero flux, is what makes the system more easily leave that phase when the magnetic flux is turned on. This suggests that the delocalization of the hole is what triggers the transition out of the ferromagnetic phase.  

The remaining 8-site geometries show $\langle h_{i} \rangle$ variability even before the magnetic flux is taken into account (see Fig.\ \ref{fig:Nagaoka_charge_8M}). It is then expected that they would have larger $\phi_0^c $ and slope values. Indeed, the critical flux for $t/U=0$ varies inversely proportional to $t_c/U$ values (see Table \ref{tab:Critical_flux}). This illustrates that to destroy kinetic ferromagnetism in the large $U$ limit with magnetic flux is highly dependent on the geometry, regardless of the number of sites.  The higher the KC variation across the cluster, the larger the  $\phi_0^c $ value (and lower $t_c/U|_{\phi=0})$ would be.

It is important to emphasize that the relationship between $\langle h_{i} \rangle$ and graph theory properties we have discussed ignores that in the presence of a magnetic flux, the arrays can no longer be regarded as undirected graphs. Interesting connections between QD arrays in the presence of Aharonov-Bohm hopping phases and graph theory would require consideration of directed graphs with nodes connected by complex weights. Directed and undirected graphs with complex weights have been applied to problems in physics, biology, and neural networks \cite{PhysRevE.109.024314,doi:10.1137/23M1584265}.  It would be an interesting challenge to explore these concepts to describe kinetic magnetism and transitions in different arrays.

\begin{table}
    \captionsetup{justification=raggedright, singlelinecheck=false}
    \centering
    \begin{tabular}{ccc}
    \hline\hline
    Geometry & $\phi_c = \pm[ (t/U) s - \phi_0^{c} ] $\\ 
    \hline
     & $s$ & $\phi_0^c$ \\
     \hline
     $2 \times 2$ & $ 19.6 $ & $0.33\pi$  \\
     $3 \times 3$ & $ 44.1 $ & $ 0.23\pi$\\
     $4 \times 4$ & $ 71.2 $ & $0.15\pi$ \\
     $5 \times 5$ & $ 136.2 $ & $ 0.11\pi$\\
     \hline
     L & $ 66.4 $ & $ 0.25\pi$  \\
     $2 \times 4$ Ladder & $49.1 $ & $ 0.23\pi$\\
     Kite & $ 26.9 $ & $0.16\pi$ \\
     2-Square & $12.6 $ & $ 0.12\pi$\\
     \hline\hline
    \end{tabular}
    \caption{$t/U$ dependence of critical flux for square and optimal arrays: geometry and $\phi^{c} (t/U) $ linear fit, in terms of slope $s$ and $\phi_0^c$.}
    \label{tab:Critical_flux}
\end{table}

\section{Discussion}

We have explored the connectivity criteria that small-scale systems must satisfy to exhibit NFM and found connections between graph theory features and the onset of NFM. Studying polaron formation in finite plaquettes through the Hubbard Hamiltonian allows us to understand the robustness of NFM and the strength of ferromagnetic clouds around the hole (Nagaoka polarons). 

In square arrays, we have found that as the number of sites increases, the critical value of hopping beyond which NFM disappears, $t_c/U$, decreases proportionally to the square of the algebraic connectivity. This shows that the hole's coherence is highly dependent on how well-connected the array is when hosting NFM.

Analyzing clusters with the same number of sites but different geometries has shown the algebraic connectivity $\lambda_2$ and the Katz centrality vector (KC) to be relevant in the resulting physical behavior. Our results point out two main conclusions when enhancing $\lambda_2$ and decreasing the variability of KC per site in an array (e.g., making the hole distribution per site as uniform as possible): 1) the ratio $t_{c}/U$ for which the transition takes place is larger, and 2) the overall strength of the Nagaoka polaron (spin correlation) decreases. The 2-Square cluster was shown to be the 8-node geometry that exhibits a more robust NFM, while the \textit{L} cluster exhibits a stronger spin correlation cloud.

We have also demonstrated that the onset of NFM is affected by an Aharonov-Bohm flux phase in finite QD array systems. For square arrays and 8-site clusters, we have studied how the NFM regime transitions to an AFM phase beyond a critical flux, exhibiting correlations similar to those obtained in the coun,ter-Nagaoaka case with next-nearest-neighbor hoppings. We have shown that a finite magnetic flux can destroy Nagaoka ferromagnetic order, and it scales as $ \sim 1/ \sqrt{A}$ in square arrays of area $A$. Our results indicate that a high variability of the hole occupancy per site $\langle  h_{i} \rangle$ in a cluster makes NFM more robust against magnetic flux when compared with arrays with the same number of sites but more homogeneous connectivity.

The geometries presented may be suitable candidates to study kinetic ferromagnetism experimentally in quantum dot plaquettes and varying magnetic fields. Such experiments could lead to a better understanding of how magnetic phases arise due to hole kinetics, and how electronic interactions in this many-electron platform affect these phenomena. It will be interesting to also explore how graph theory relates to kinetic magnetism when a magnetic flux is considered. This would need consideration of directed graphs with complex weights, which may in turn allow connections with other systems.

%\nocite{*}

\bibliography{bibliography.bib}% Produces the bibliography via BibTeX.

%\newpage
%
\appendix

% Redefine the name within the appendix scope
\renewcommand{\figurename}{FIG. A}
% Reset the figure counter so it starts from 1 again (optional, depending on your needs)
\setcounter{figure}{0}

\section{Algebraic connectivity} \label{Appendix}

\begin{figure}[h]
    \includegraphics[width=0.95\linewidth]{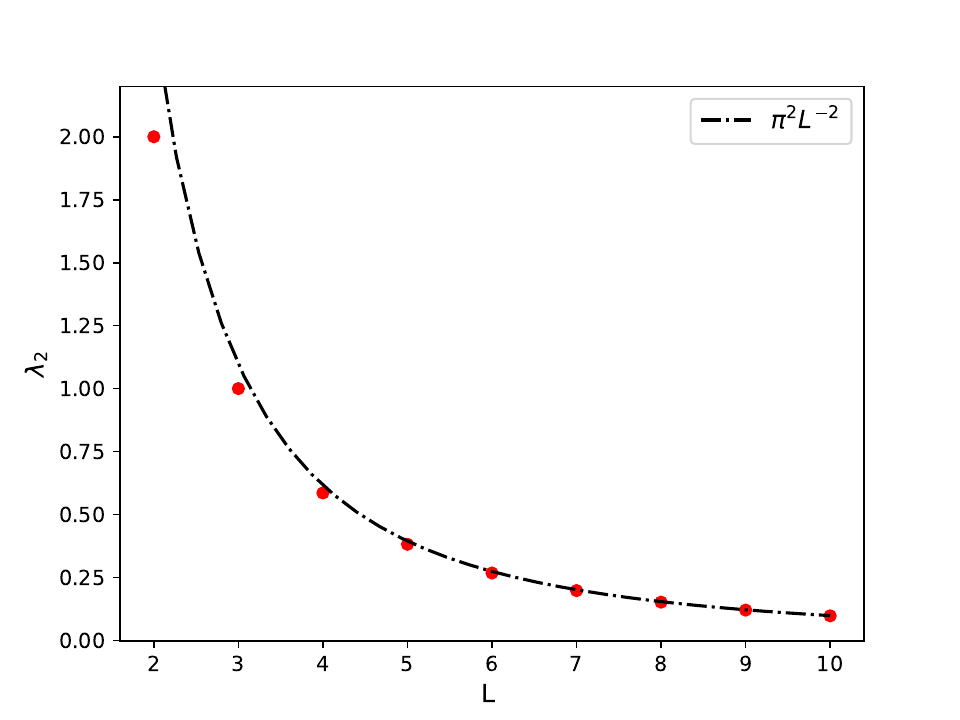}
    \centering
    \caption{\justifying Algebraic connectivity as a function of $L$ in square arrays of $L \times L$ is found to be $\lambda_2  \simeq \pi^2 /L^2$. See text.}
    \label{fig:lambda2}
\end{figure}

Algebraic connectivity scales as $L^{-2} $ in square arrays, as shown in Fig.\ A\ref{fig:lambda2}. The same relationship holds for ladders of $ 2 \times L$. This shows that the well-connectedness of a graph depends on its largest dimension \cite{fiedler1989laplacian}.

\renewcommand{\figurename}{FIG. B}
% Reset the figure counter so it starts from 1 again (optional, depending on your needs)
\setcounter{figure}{0}
\renewcommand{\tablename}{TABLE B}
\setcounter{table}{0}

\section{Ten-site geometries} \label{AppendixB}

\begin{figure*}[h]
    \centering
    \begin{subfigure}[b]{0.25\textwidth}    
        \begin{overpic}[width=\linewidth,grid=false]{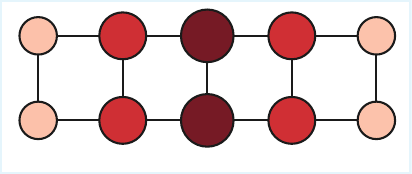}
            % coordinates: (x,y) in percent, (0,0)=bottom-left, (100,100)=top-right
            \put(2,77){\textbf{(a)}} % label in top-left
        \end{overpic}          
        
        \label{fig:g1}
    \end{subfigure}
    \hfill
    \begin{subfigure}[b]{0.25\textwidth}
        \begin{overpic}[width=\linewidth,grid=false]{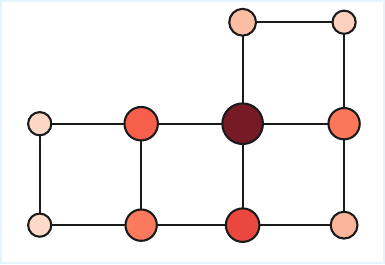}
            % coordinates: (x,y) in percent, (0,0)=bottom-left, (100,100)=top-right
            \put(2,70){\textbf{(b)}} % label in top-left
        \end{overpic}       
        \label{fig:g2}
        
    \end{subfigure}
    \hfill
    \begin{subfigure}[b]{0.25\textwidth}
        \begin{overpic}[width=\linewidth,grid=false]{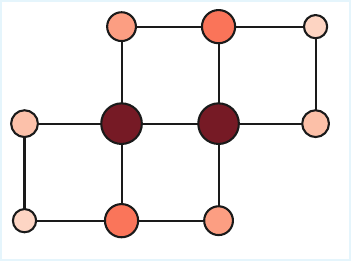}
            % coordinates: (x,y) in percent, (0,0)=bottom-left, (100,100)=top-right
            \put(2,70){\textbf{(c)}} % label in top-left
        \end{overpic}       
        \label{fig:g5}
    \end{subfigure}
    \hfill
    \vspace*{1cm}
    \hspace*{1cm}
    \begin{subfigure}[b]{0.25\textwidth}
        \begin{overpic}[width=\linewidth,grid=false]{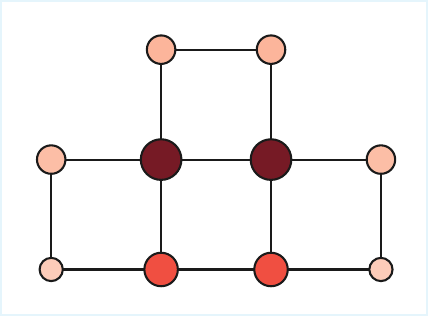}
            % coordinates: (x,y) in percent, (0,0)=bottom-left, (100,100)=top-right
            \put(2,68){\textbf{(d)}} % label in top-left
        \end{overpic}       
        \label{fig:g3}
    \end{subfigure}
    %\hfill
    \hspace*{1cm}
    \begin{subfigure}[t]{0.25\textwidth}
        \begin{overpic}[width=0.8\linewidth,grid=false, angle=-90]{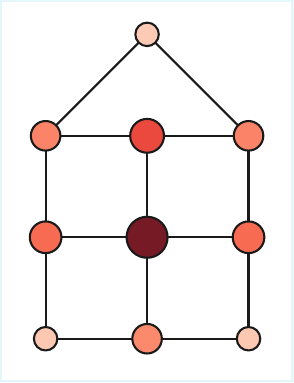}
             %coordinates: (x,y) in percent, (0,0)=bottom-left, (100,100)=top-right
            \put(1.,74){\textbf{(e)}} % label in top-left
        \end{overpic}       
        \label{fig:g2}
    \end{subfigure}
    \hfill
    %\vspace*{0.5cm}
    %\hspace*{5cm}
    \begin{subfigure}[b]{0.5\textwidth}
        \begin{overpic}[width=0.7\linewidth,grid=false]{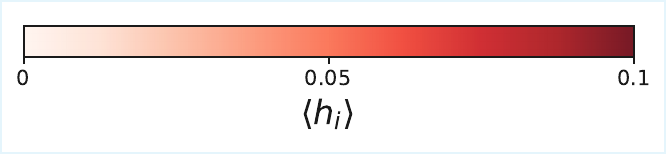}
             %coordinates: (x,y) in percent, (0,0)=bottom-left, (100,100)=top-right
            %\put(2,48){\textbf{(e)}} % label in top-left
        \end{overpic}       
        \label{fig:bar}
    \end{subfigure}
    \centering \caption{\justifying Hole distribution per site in the proposed clusters with ten sites at $t<t_c$. The color of each circle conveys $\langle  h_{i} \rangle$, while its diameter is proportional to the Katz centrality (KC). Notice that in the \textit{b-array}, the hole is mostly concentrated in the bulk-site (larger diameter), whereas in the \textit{a-array}, the hole is distributed over more sites. In the \textit{e-array}, the hole is distributed more evenly among sites, corresponding to a low fluctuation in its KC (site's diameter). }
    \label{fig:hole_KC_10sites}
\end{figure*}

Geometries of ten sites that host NFM and satisfy the connectivity conditions described in the main text. In Table B\ref{tab:geometries_10sites}, we report their corresponding $t_c/U$ and $\lambda_2$ values. Figure B \ref{fig:hole_KC_10sites} shows hole distribution (represented by color red) and KC (given by the diameter of the circles) in these arrays.

\begin{table}[h]
    \captionsetup{justification=raggedright, singlelinecheck=false}
    \centering
    \begin{tabular}{lclccl}
    \hline\hline
     Array &  $t_{c}/U$ & & $\lambda_2$  \\
     \hline
      a &  0.007 && 0.38 \\
     b &  0.007  && 0.46  \\
     c &    0.008  & &  0.53 \\
     d &    0.009  & &0.62 \\
     e &    0.0125  && 0.795 \\
     \hline\hline
    \end{tabular}
    \caption{Proposed arrays with 10 sites: Array label, transition point and algebraic connectivity $\lambda_2$.}
    \label{tab:geometries_10sites}
\end{table}

%\newpage

\end{document}